\documentclass{iopart}
\usepackage{epsfig}
\usepackage{lscape}
\usepackage{amsfonts} 
\newcommand{\av}[1]{\left\langle{#1}\right\rangle}

\begin{document}

\title{Current-voltage correlations in interferometers} 
\author{Heidi F\"orster\dag, Peter Samuelsson\ddag, Markus B\"uttiker\dag}
\address{\dag\ D\'epartement de Physique Th\'eorique, Universit\'e de Gen\`eve, CH-1211 Gen\`eve 4,
Switzerland} 
\address{\ddag\ Departement of Solid State Theory, Lund University, S\"olvegatan 14 A, S-223 62 Lund, Sweden} 
\date{\today}

\begin{abstract}
We investigate correlations of current at contacts and voltage fluctuations at
voltage probes  coupled to interferometers.  The results are compared with
correlations of  
current and occupation number fluctuations at dephasing probes. We use a 
quantum Langevin approach for the average quantities and their fluctuations.
For higher order correlations we develop a stochastic path integral 
approach and find the generating functions of voltage or occupation number
fluctuations. We also derive a generating function for the joint distribution
of voltage or occupation number at the probe and current fluctuations at a
terminal of a conductor. For energy independent scattering we found earlier
that the generating function of current cumulants in interferometers with a
one-channel dephasing  or voltage probe are identical. Nevertheless, the
distribution function for  voltage and the distribution function for
occupation number fluctuations  differ, the latter being broader than that of
former in all examples considered here.  
\end{abstract}
\maketitle 

\section{Introduction.}

The quantization of charge and the diffraction of quantum waves lead to
fluctuations in the transport state of electrical conductors even in the limit
of absolute zero temperature \cite{butrew}.  
These fluctuations provide additional information not available from
conductance measurements alone. This has led to considerable theoretical
\cite{lll,nazarov,SPI1,nagaev} and experimental efforts
\cite{reulet,bomze,gustavsson,fujisawa,timofeev} to characterize fluctuations
not only on the level of noise but including all higher order  
cumulants. These efforts have been focused almost exclusively on the
statistics of transferred charge \cite{lll,nazarov,SPI1}, i.e.~on the cumulants
of currents measured at a contact of a conductor. Clearly, there are however
also other quantities of interest. For example, fluctuations of the charge
inside a conductor \cite{pilgram03,SPI2,utsumi} (as opposed to transferred
charge) are closely connected to electrostatic potential fluctuations in a
conductor \cite{seelig03} and are thus essential in determining the long range
Coulomb interaction of two nearby conductors.  Still another example is
fluctuations of voltage. The statistics of voltage have been investigated  for
a mesoscopic conductor in series with a classical  impedance
\cite{kindermann03,hakonen} and for networks of two-terminal
conductors \cite{kindermann04}.

In this work we consider voltage probes which connect to a
conductor within its phase-coherent volume \cite{mb88}. We determine the
fluctuations of the voltage at the probe \cite{Been92,texier,oberholzer2} and
determine the joint probability distribution between the voltage fluctuations
and the charge transferred into a contact of the conductor.  Voltage probes are
real elements of mesoscopic conductors and are used to gain information of the
interior state of the system. We compare the fluctuations at a voltage probe
with the fluctuations at a dephasing probe. This latter probe conserves not
only charge but also energy \cite{datta,Jong96}. At a dephasing probe the
occupation 
number fluctuates \cite{texier}.  Voltage and dephasing probes provide a
simple means to introduce incoherent events in an otherwise quantum coherent
conductor \cite{mb88,Jong96}. 

We examine these correlations with the help of a (quantum) Langevin approach
\cite{mb90,mb92}. In a second step we derive a generating function for the
correlations using a semi-classical path integral approach
\cite{short,long}. On the level of conductance and noise the Langevin approach
and the saddle point solution of the path integral approach are shown to
agree. We have earlier demonstrated that for the statistics of transferred  
charge 
both a single channel voltage probe and the single channel dephasing probe
lead to the same counting functional if the scattering matrix of the conductor
is energy independent in the range of applied voltages
\cite{short,long}. Interestingly, we find that this equivalence does not hold
for the joint distribution of voltage (or occupation number) fluctuations and
transferred charge.  

The characterization of the transport fluctuations not only in terms of
transferred charges but also of internal properties is a central motivation of
this work. Voltage fluctuations at a probe test carrier fluctuations that can
still be evaluated in an approach that considers samples with terminals
only. It is particularly interesting for conductors which exhibit quantum
interference since voltage and dephasing probes are phase breaking elements.  
A carrier that enters a probe is eventually replaced by a carrier that is
re-emitted into the conductor with a phase (and in the case of a voltage probe,
an energy) that is unrelated to the exiting carrier. This carrier is detected
in the voltage probe and thus does not contribute to certain interference
processes.  Phase breaking and which path detection are
generally closely linked phenomena \cite{stern}. This provides
another motivation to investigate the fluctuations at the probes (the which
path detectors) and their correlations with the transmitted current.

\section{Correlations: Langevin approach.}\label{Langevin}

The Langevin approach is a convenient tool to
investigate current and voltage correlations in mesoscopic conductors
\cite{mb90,mb92}.  This has been demonstrated in numerous
papers, see e.g. Refs.~\cite{butrew, Been92, texier, brouwer,
  mb, clerk, chung06}. Importantly, the Langevin approach 
also allows one to obtain information on the fluctuation properties of a
probe connected to the mesoscopic conductor.
In this section we investigate the average current and potentials as well as
the auto- and cross-correlations. It is possible to extend the Langevin
approach successively to higher order correlations, giving rise to 
cascade corrections \cite{nagaev}. In section \ref{FCS} we instead
directly determine the generating function of the correlations,
making use of a stochastic path integral approach  \cite{SPI1}.

We consider a multi terminal quantum coherent conductor first without being
connected to a probe. 
The current through the conductor fluctuates. When the conductor is embedded
in  a zero-impedance external circuit, all contacts are characterized by
Fermi distribution functions which are constant in time. Only
bare current fluctuations take place due to scattering in the conductor,
described by a unitary scattering matrix $\mathcal{S}$. 
The total current in contact $\alpha$ can be expressed as a sum of
the average current and the current fluctuations \cite{mb90,mb92}
\begin{equation}\label{current}
	I_\alpha=\av{I_\alpha}+\delta I_\alpha=\sum_\beta G_{\alpha\beta}
	V_\beta+\delta I_\alpha, 
\end{equation}
where $V_\beta$ is a voltage applied to reservoir $\beta$, and  
$G_{\alpha\beta}$ the conductance matrix
\begin{equation}\label{G}
	G_{\alpha\beta}=\frac{e^2}{h}\int dE \left(-\frac{\partial 
	f}{\partial
	E}\right)[N_\alpha\delta_{\alpha\beta}-T_{\alpha\beta}]. 
\end{equation}
The equilibrium Fermi function is called $f$, the number of channels in lead
$\alpha$ is $N_\alpha$, and
$T_{\alpha\beta}=T_{\alpha\beta}(E)= 
Tr[\mathcal{S}^\dagger_{\alpha\beta}(E)\mathcal{S}_{\alpha\beta}(E)]$
denotes the total probability that a particle is transmitted from terminal
$\beta$ to terminal $\alpha$ where the trace runs over the transport channels.

Like the conductances, the noise correlations can be expressed in terms of
scattering matrices. The
zero frequency noise and correlations are $C_{\alpha\beta}=\av{\delta 
  I_\alpha \delta I_\beta}$.  Note that the  definition of the noise as
given for example in Ref.~\cite{butrew} usually differs by a factor of $2$. 
The noise can be expressed as
$C_{\alpha\beta}=C_{\alpha\beta}^{eq}+C_{\alpha\beta}^{tr}$, a sum of an
equilibrium contribution $C_{\alpha\beta}^{eq}$ stemming 
from noisy incident beams in one contact proportional to
$f_\alpha(1-f_\alpha)$,  and a transport contribution $C_{\alpha\beta}^{tr}$
which depends on Fermi functions of two different reservoirs
$f_\gamma(1-f_\delta)$ (with $\gamma \not= \delta$). Following
Ref.~\cite{mb92} we find
\begin{eqnarray}
	C_{\alpha\beta}^{eq}&=&
	 \frac{e^2}{h}\int dE
	 (2\delta_{\alpha\beta}N_\alpha f_\alpha(1-f_\alpha)-\nonumber\\
	 &&\hspace{13mm}T_{\beta\alpha}f_\alpha(1-f_\alpha)-  
	T_{\alpha\beta}f_\beta(1-f_\beta)),\label{S1}\\  
	C_{\alpha\beta}^{tr}&=& 
	\frac{e^2}{2h}\int dE \sum_{\gamma,\delta}
	Tr[\mathcal{S}_{\alpha\gamma}^\dagger
	 \mathcal{S}_{\alpha\delta}\mathcal{S}_{\beta\delta}^\dagger  
	\mathcal{S}_{\beta\gamma}]\nonumber\\
	&&\hspace{13mm}
	 \left(f_\gamma(1-f_\delta)+f_\delta(1-f_\gamma)-
	2f_\alpha(1-f_\alpha)\right).   \label{S2}  
\end{eqnarray}
In the case when temperature is negligible compared to the applied bias, the
contribution $C_{\alpha\beta}^{eq}$ vanishes and  the term
$C_{\alpha\beta}^{tr}$ represents the pure shot noise.

\subsection{Voltage probe.}\label{langevolt}
A voltage probe is a  large metallic contact connected to the conductor. The 
potential $V_p$ at the contact is left floating and there is no current drawn
at the probe.  The potential $V_p=V_p(t)$ exhibits fluctuations on the time
scale $\tau_d$, which is given by the $RC$-time of a classical circuit.  The
 fluctuations originate from the response of the potential to the
injected charges: Incoming charges make the potential $V_p$ rise. This leads
in turn to an increase in the outgoing current which reduces the potential
again and so on. 
This mechanism is formally expressed by vanishing current  and current
fluctuations at the probe for frequencies smaller than $1/\tau_d$:
\begin{equation}\label{probecondition}
	\av{I_p} =0, \hspace{2cm}\Delta I_p=0.
\end{equation}
The voltage probe is described by an equilibrium Fermi function $f_p(V_p)$,
where thermalization by inelastic scattering is assumed to be much faster
than the delay time $\tau_d$, and where the temperature is determined by the
lattice temperature. Therefore, a particle scattering via the voltage probe
looses not only its phase memory but also changes its energy.

The potential fluctuations at the probe give rise to  additional
fluctuations of the current in the terminals \cite{Been92, texier}. The
potential at the probe can be separated into a 
constant and a fluctuating part, $V_p=\bar V_p+\Delta V_p$. 
Following Eq.~(\ref{current}) the average current $\av{I_\alpha}$ and the
total fluctuations $\Delta I_\alpha$ are in linear response given by
\begin{eqnarray}
	\av{I_\alpha} &=&\sum_{\beta\not=
 	p}G_{\alpha\beta}V_\beta+G_{\alpha p} \bar V_p,\label{Ialpha}\\
	\Delta I_\alpha&=&G_{\alpha p} \Delta V_p +\delta
 	I_\alpha.\label{DeltaI} 
\end{eqnarray}
Both the constant value $\bar V_p$ and the fluctuating part $\Delta V_p$  of
the potential at the probe 
are determined by the condition Eq.~(\ref{probecondition}). Throughout this
paper we consider that only one of the contacts (taken to be $\alpha=1$) has
an elevated potential $eV$ and all other current terminals are
grounded. Solving Eqs.~(\ref{Ialpha}) and (\ref{DeltaI}) with $\alpha=p$, the
potential at the probe is found to be $	\bar V_p=-G_{p1}V/G_{pp}$, and $
	\Delta V_p = -\delta I_p/G_{pp}$.
Using this we find for the  auto-correlations of the potential and the cross
correlations of the potential with the current in terminal $\alpha$
\begin{eqnarray}
	\av{\Delta V_p^2}&=& \av{\delta
	I_p^2}/G_{pp}^2=\frac{C_{pp}}{G_{pp}^2},\label{SVV}\\ 
	\av{\Delta V_p\Delta I_\alpha}&=& \left(\frac{G_{\alpha
	p}}{G_{pp}}\av{\delta I_p^2}-\av{\delta I_p \delta I_\alpha}
	\right)/G_{pp}=\frac{G_{\alpha
	p}C_{pp}}{G_{pp}^2}-\frac{C_{p\alpha}}{G_{pp}}. \label{CVbeta}
\end{eqnarray}
These expressions can be evaluated for a given scattering matrix using
Eqs.~(\ref{G}), (\ref{S1}) and (\ref{S2}), where only the constant Fermi
functions enter, i.e.~$f_p=f_p(\bar V_p)$.  For higher order correlations the  
fluctuations of the Fermi functions have to be taken into account. As pointed
out above, this leads to  
noise of noise and results in a cascade corrections similar to the one known
from the quasi-classical fluctuating Boltzmann equation \cite{nagaev}.

\subsection {Dephasing probe.}\label{langedeph}
\begin{figure}[b]
\centerline{\psfig{figure=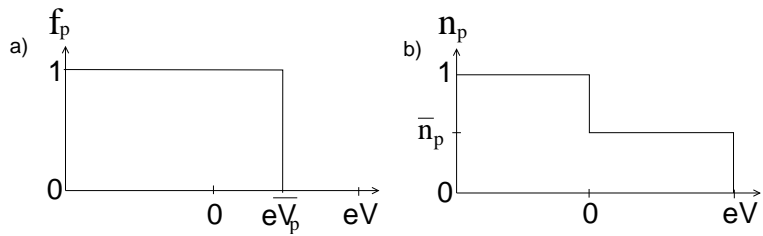,width=10.0cm}}
\caption{a) The Fermi distribution function of the voltage probe takes on only
the values zero or one (at zero temperature). The position of the step is
given by the potential at the probe. b) The
distribution function of the dephasing probe is a two-step function and has a
value between zero and one in the interval $[0,eV]$.} 
\label{distribution}
\end{figure}
Dephasing probes are conceptual tools to
describe quasi-elastic dephasing \cite{datta,Jong96}. In the case of a
dephasing 
probe,  a particle entering the probe is incoherently re-emitted within 
the same energy interval $[E,E+dE]$. It looses its phase, but the change in
energy $dE$ is much smaller than the external bias or the temperature.  
Because scattering in each energy interval is independent, the distribution
function $n_p(E)$ is -unlike the Fermi function describing the voltage probe-
a highly non-equilibrium distribution function. It shows
independent fluctuations in 
each energy interval, $n_p(E)=n_p(E,t)$, which occur on the time scale
$\tau_d$, the delay time of the probe.  
In a dephasing probe, the current per energy interval $i_p(E)$ and its
fluctuations are conserved up to the frequency $1/\tau_d$:
\begin{equation}\label{dephcondition}
	\av{i_p} =0, \hspace{2cm}\Delta i_p=0.
\end{equation}
Similar to the potential of the voltage probe, the non-equilibrium
distribution function is written as a constant and 
a fluctuating part $f_p \equiv \bar n_p+\Delta n_p$. All other terminals
are still 
characterized by constant equilibrium Fermi functions. The average current per
energy $\av{i_\alpha(E)}=\av{i_\alpha}$ and the total current fluctuations per
energy $\Delta i_\alpha(E)=\Delta i_\alpha$ are given by
\begin{eqnarray}
	\av{i_\alpha}&=&\sum_{\beta \not= p} g_{\alpha\beta}
	f_\beta+g_{\alpha p} \bar n_p,\\
	\Delta i_\alpha &=& g_{\alpha p}\Delta n_p +\delta i_\alpha.
\end{eqnarray}
In analogy to the expressions for the energy integrated current
$I_\alpha=\int dE i_\alpha(E)$, an energy dependent conductance matrix
$g_{\alpha\beta}=g_{\alpha\beta}(E)$ can be defined, 
\begin{equation}
	g_{\alpha\beta}=\frac{e}{h}[N_\alpha\delta_{\alpha\beta}-T_{\alpha
	\beta }].
\end{equation}
Also for the zero-frequency noise we find an energy dependent function from 
$C_{\alpha\beta}=\int dE c_{\alpha\beta}(E)$.
Again the function $c_{\alpha\beta}=c_{\alpha\beta}(E)=\av{\delta
  i_\alpha \delta  i_\beta} $ splits into an equilibrium and a transport
contribution 
\begin{equation}
	c_{\alpha\beta}=c_{\alpha\beta}^{eq}+c_{\alpha\beta}^{tr},
\end{equation}
where
$c_{\alpha\beta}^{eq}$ and $c_{\alpha\beta}^{tr}$ are the integrands of
Eqs.~(\ref{S1}) and (\ref{S2}) respectively. $c_{\alpha\beta}^{eq}$ contains
the noise due to a noisy incident beam in one contact, while
$c_{\alpha\beta}^{tr}$ involves particles from two different contacts. 

Using Eq.~(\ref{dephcondition})  the noise and correlations at the probe
are  \cite{texier, vanLangen}
\begin{eqnarray}
	\av{\Delta n_p^2}&=& \av{\delta
	i_p^2}/g_{pp}^2=\frac{c_{pp}}{g_{pp}^2},\label{Snn}\\ 
	\av{\Delta n_p\Delta i_\alpha}&=& \left(\frac{g_{\alpha
	p}}{g_{pp}}\av{\delta i_p^2}-\av{\delta i_p \delta i_\alpha}
	\right)/g_{pp}=\frac{g_{\alpha
	p}c_{pp}}{g_{pp}^2}-\frac{c_{p\alpha}}{g_{pp}}.\label{Cnbeta} 
\end{eqnarray}
Here we can see the first consequence of the different electron distributions
(see Fig.~\ref{distribution})
in the voltage and dephasing probe. Although Eqs.~(\ref{SVV})
and (\ref{Snn}) have exactly the same structure, the
evaluation leads to different results. The reason is that in the case of a
voltage probe the contribution $C_{pp}^{eq}$ to the noise vanishes at
zero temperature, while for a dephasing probe, the corresponding expression
$c_{pp}^{eq}$ is non-zero even at zero temperature since the factor
$\bar n_p(1-\bar n_p)$ is finite in the energy interval $[0,eV]$ of interest. 
Interestingly, in Eqs.~(\ref{CVbeta}) and (\ref{Cnbeta}), inserting the
expressions for the conductance and  
correlators from Eqs.~(\ref{G})-(\ref{S2}), it follows that the terms
proportional to  $\bar n_p(1-\bar n_p)$ drop out. Consequently, for energy
independent scattering, the  
energy integrated correlations $\av{\Delta V_p\Delta I_\alpha}/V$ equal the
correlations  $\av{\Delta n_p\Delta i_\alpha}$.

\subsection{Correlations and interference.}\label{langeex}
It is interesting to investigate how and if 
 interference appears  in the fluctuations of the potential and the occupation
 number of the probe or in the correlations with the current in
 one of the leads. To this end we evaluated Eqs.~(\ref{SVV}), (\ref{CVbeta})
and (\ref{Snn}), (\ref{Cnbeta}) for four different setups: a beam splitter and
 three 
 interfering systems, the Mach-Zehnder interferometer, the double barrier and
 the triple barrier. 

These examples also illustrate an interesting feature of the different nature
of voltage and dephasing probes. It was shown in Refs.~\cite{short, long} that
the auto- and cross  correlations (actually all cumulants) of the current in a
conductor coupled to a single mode probe are independent of whether the probe
is a voltage or a dephasing probe for the case that scattering is 
energy  independent. Interestingly, this does not hold for the fluctuations
of the probe: there is a clear difference between the potential fluctuations
of a voltage probe and the occupation number fluctuations of the dephasing
probe. 

\begin{figure}[b]
\centerline{\psfig{figure=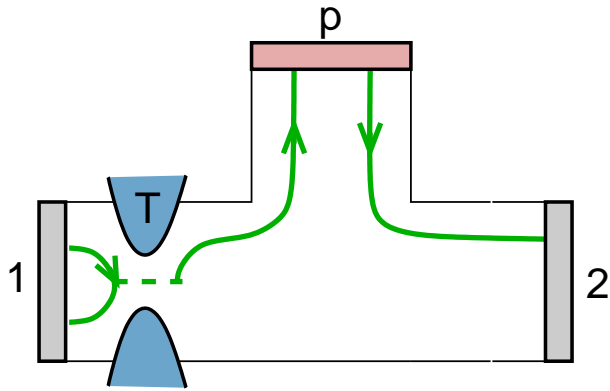,width=8cm}}
\caption{A quantum point contact in the quantum Hall regime forms a beam
  splitter with transmission probability $T$. A perfectly coupled probe is
  attached along the upper edge.}  \label{beamsplitter}
\end{figure}

All the examples are quantum Hall bars of different complexity subject to a
high magnetic field. Current is
transported by an edge state at filling factor one, where carriers move in
one specific direction along the edge of the structure.  A 
quantum point contact in such a structure forms a beam
splitter: a particle is either transmitted and continues along the same edge,
or it is reflected and moves into the opposite direction along the opposite
edge.  
The beam splitter is the smallest unit that the interfering structures are
composed of, we chose it here as a simple and instructive example to show how a
voltage or dephasing probe behaves. For simplicity we consider the scattering
matrix of all structures to be energy independent. 
In the following we present briefly the setups of the four examples. The
results  are collected in table \ref{table1} and table \ref{table2}.

\begin{itemize}

\item The {\bf beam splitter} \cite{oberholzer1,Oliver99} is formed by a
  quantum   point contact with 
  transmission probability $T$, situated between two terminals $1$ and $2$, as
  shown in Fig.~\ref{beamsplitter}. A probe -either a voltage or a dephasing
  probe- is
perfectly coupled to the setup, such that every particle transmitted through
  the beam 
splitter enters the probe and moves then from the probe into contact $2$.

\item An electronic  {\bf Mach-Zehnder interferometer} (MZI) is shown
  in Fig.~\ref{mzifig}.  
\begin{figure}
\centerline{\psfig{figure=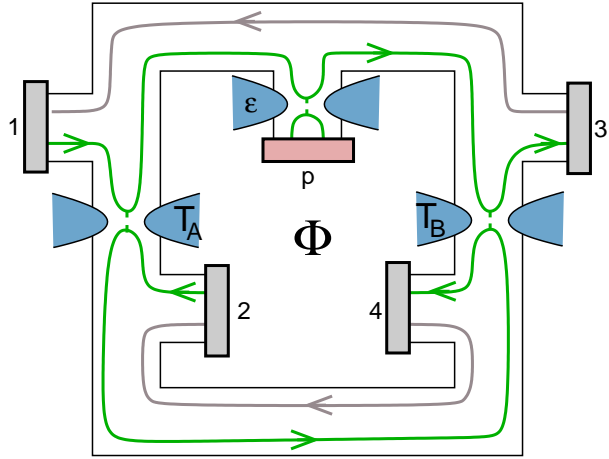,width=8.0cm}}
\caption{A Mach-Zehnder interferometer with its upper arm
coupled to a dephasing or voltage probe with coupling strength $\varepsilon$.
An Aharonov-Bohm flux generates a phase $\Phi$.}
\label{mzifig}
\end{figure}
It consists of two arms connected to four electronic
reservoirs $1$ to $4$ via beam splitters $A$ and $B$. The
transmission (reflection) probabilities of the beam splitters are $T_A$ and
$T_B$ ($R_A$ and $R_B$) respectively.
Interference occurs, because the electrons have two alternative 
paths to propagate through the interferometer between beam
splitter $A$ and $B$. An
Aharonov-Bohm flux threads the two arms, and the different vector
potentials in 
the two arms lead to a phase difference $\Phi$. This
difference creates a characteristic flux-periodicity in the 
interference pattern, the Aharonov-Bohm effect. 
A voltage or dephasing probe is attached to the upper arm. Particles entering
the probe loose their phase and, therefore, the interference is reduced,
controlled by the coupling strength $\varepsilon$.
Thus the Mach-Zehnder interferometer is useful as a conceptually simple
interferometer and has been used in theoretical discussions of dephasing
\cite{Seelig,Marq1,Marq2,Forster05,Chung}.  
Such a setup was recently realized experimentally \cite{Ji,strunk,neder} and
has since generated further works
\cite{neder07,Sukh06,expmarquardt,marqneder}.  If the length of the two
interferometer arms is not equal, the transmission through the interferometer
will be energy dependent due to the geometric phase a particle acquires. For
finite temperature, interference effects
will be averaged out. In order to determine purely
the effect of dephasing and 
voltage probes, we concentrate here on the ideal case of an equal arm lengths
interferometer.

\item The {\bf double barrier}, shown in Fig.~\ref{doublebarrier}, is a two
terminal conductor with two beam splitters \cite{vwees,alphen}.  
In the region between the barriers
(the dot) a particle can perform multiple loops and undergo resonant
scattering. For each loop in the dot, the electron picks up a phase $\Phi_d$.
We consider zero temperature and an energy independent phase.  The
interference is here not a consequence of two spatially distinct paths but of
multiple possible paths through the closed orbit inside the dot.  
A probe is coupled to the dot with coupling strength
$\varepsilon$. Since particles that scatter via the probe loose their phase,
coherence is partially destroyed. For perfect coupling, $\varepsilon=1$, all
particles entering the dot pass on into the probe, then the system is
  incoherent   and in the sequential tunneling regime.
\begin{figure}[b]
\centerline{\psfig{figure=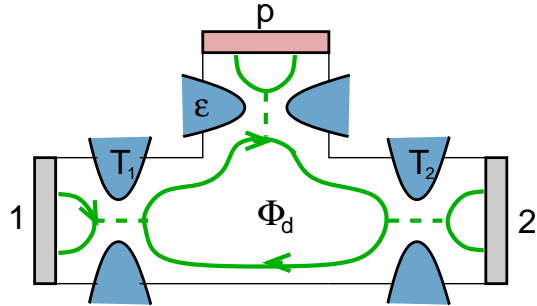,width=7.0cm}}
\caption{The double barrier quantum Hall interferometer with a resonant
  state. A probe couples to the resonant state with coupling 
  strength $\varepsilon$. $\Phi_d$ is the total phase acquired in one loop. } 
\label{doublebarrier}
\end{figure}

\item In the {\bf triple barrier} setup, an additional barrier
  in series is  inserted into the double barrier setup, see
  Fig.~\ref{triplebarrier}.  Now there are two dots \cite{vvaart, kiesslich,
  jakobs} with   two   different phases   $\Phi_L$  
and $\Phi_R$. Resonances can occur in either one of the dots or through both
dots simultaneously, rendering the scattering matrix considerably more
complex. The limiting case $R_2 \rightarrow 0$ however represents the double
barrier with $\Phi_L+\Phi_R=\Phi_d$. We consider here a probe coupled to
the left dot, thus only the interference in the left dot is destroyed by the
probe.
\begin{figure}[b]
\centerline{\psfig{figure=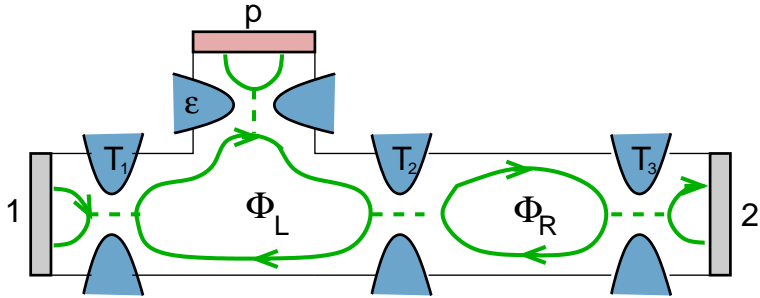,width=10.0cm}}
\caption{The triple barrier with two resonant states. A single probe is
 coupled to the  left resonant state. } 
\label{triplebarrier}
\end{figure}

\end{itemize}


\begin{table}
\caption{The mean squared fluctuations of the potential and the occupation
  number for the four examples. The
  transmission probabilities of all point contacts in the double and triple
  barrier structures are 
  equal, $T=1-R$. Interference effects are present only in the latter two.  } 
  \label{table1}  
\begin{center}
\begin{tabular}{|p{0.2\linewidth}|p{0.34\linewidth}|p{0.34\linewidth}|}   
\hline      
 & 
	\begin{center}$\av{\Delta V_p^2}\frac{e}{hV}$ \end{center}&
	\begin{center}$\av{\Delta n_p^2}\frac{1}{h}$\end{center} \\
\hline\hline
	Beam splitter &
	\begin{center}$T(1-T)$\end{center} & \begin{center}$2
	T(1-T)$\end{center} \\
\hline
	Mach-Zehnder interferometer &
	\begin{center}$\frac{2-\varepsilon}{\varepsilon}T_A(1-T_A)$\end{center}
	&  
	\begin{center}$\frac{2}{\varepsilon}T_A(1-T_A)$\end{center} \\
\hline
	Double barrier &
	\begin{center}$\frac{R((1+R^2)(2-\varepsilon)+4R\sqrt{1-\varepsilon}\cos\Phi_d)} 
	{\varepsilon(1-R)(1+R)^3}$ \end{center}&
	\begin{center}$\frac{2 R(1+R^2(1-\varepsilon)+2R\sqrt{1-\varepsilon}\cos\Phi_d)}
	{\varepsilon(1-R)(1+R)^3}$ \end{center}\\
\hline
 	Triple barrier &
	\begin{center}$\frac{R}{\varepsilon}\mathcal{F}(R,\Phi_R)
	[(1+3R^2)(2-\varepsilon)+4R(1+R)\sqrt{1-\varepsilon}\cos\Phi_L+
	4R^2\sqrt{1-\varepsilon}\cos(\Phi_L-\Phi_R)+
	2R(1+R)(2-\varepsilon)\cos\Phi_R+ 
	4R\sqrt{1-\varepsilon}\cos(\Phi_L+\Phi_R)]$\end{center}&
	\begin{center}$\frac{2R}{\varepsilon}\mathcal{F}(R,\Phi_R)
	[1+3R^2-2R^2\varepsilon+2R(1+R)\sqrt{1-\varepsilon}\cos\Phi_L+
	2R^2\sqrt{1-\varepsilon}\cos(\Phi_L-\Phi_R)+
	2R(1+R(1-\varepsilon))\cos\Phi_R+ 
	2R\sqrt{1-\varepsilon}\cos(\Phi_L+\Phi_R)]$\end{center}\\ 
	&with \begin{center}$\mathcal{F}(R,\Phi_R)=\frac{1+R^2+2R\cos\Phi_R}
	{(1+R+2R\cos\Phi_R)^3}$\end{center} & \\
\hline
\end{tabular}
\end{center}
\end{table}
\begin{table}
\caption{Current voltage correlations $\av{\Delta V_p
  \Delta I_\alpha}$ and current occupation number fluctuations $\av{\Delta n_p
  \Delta i_\alpha}$. For energy independent scattering these two correlations
  are identical. }    \label{table2}  
\begin{center}
\begin{tabular}{|p{0.2\linewidth}|p{0.36\linewidth}|}    
\hline      
 & 
	\begin{center}$\av{\Delta V_p\Delta I_\alpha}\frac{1}{eV}=\av{\Delta
 n_p\Delta i_\alpha}\frac{1}{e}$ 
	\end{center} 
	\\  
\hline\hline
	Beam splitter &
	\begin{center}$\pm T(1-T)$ \mbox{for $\alpha=1$ (+) and $\alpha=2$ (-)}
	\end{center}\\ 
\hline
	Mach-Zehnder interferometer &
	\begin{center}$T_A(1-T_A)(1-2T_B)$ for $\alpha=3$\end{center} \\
\hline
	Double barrier &
	\begin{center}$-\frac{R(1-R)}{(1+R)^3}$ for $\alpha=2$\end{center}\\
\hline
	Triple barrier &
	\begin{center}$R(1-R)\mathcal{F}(R,\Phi_R)[-1+3R+2R\cos\Phi_R]$ for
	$\alpha=2$ 
	\end{center}\\ 
\hline
\end{tabular}
\end{center}
\end{table}


In table \ref{table1} the results for the auto-correlations of the voltage
  $\av{\Delta   V_p^2}$ and of the occupation number 
$\av{\Delta n_p^2}$ are listed. The different electron distributions of
voltage and dephasing probes,  as pointed out already in section
  \ref{langedeph}, become apparent: the fluctuations of the occupation number
  $\bar n_p (E)$ in a dephasing probe 
are larger than the potential fluctuations of the voltage probe.  This is most
easily seen in the example of the 
beam splitter, where the fluctuations of $\bar n_p (E)$ are
twice as large as the fluctuations of $\bar V_p$. Note that occupation numbers
at different energies  
are uncorrelated but that the fluctuation spectrum, in the limit of zero
temperature, and in the limit that the energy dependence of the scattering
matrix can be neglected, is independent of energy for $0 \le E \le eV$.

For finite coupling, the
fluctuations are proportional to $1/\varepsilon$ and diverge for $\varepsilon
\rightarrow 0$ for both voltage and dephasing 
probe. This means the fewer particles 
enter the probe, the stronger are the fluctuations. 
Note that the differences in the fluctuations of voltage and dephasing probes
are of order one compared to $1/\varepsilon$ and, therefore, become less
important for small $\varepsilon$. 
(For the beam splitter, the table includes only 
the strong coupling result $\varepsilon = 1$. The above statement applies also 
to the beam splitter with a weakly coupled probe). 

The noise of the probe in the MZI is independent of the phase
$\Phi$. This is not 
surprising since particles enter the probe before they could interfere. 
This is in accordance with Ref.~\cite{Seelig} which demonstrates that 
charge density fluctuations in an MZI arm are independent of phase. 
In the double barrier, particles
enter the probe after performing multiple loops inside the dot  
and thus the fluctuations depend on the phase $\Phi_d$ picked up.
In the triple barrier different paths containing different resonant passages
are possible and the noise contains a combination of $\Phi_L$, $\Phi_R$,
$\Phi_L+\Phi_R$ and $\Phi_L-\Phi_R$.

Table \ref{table2} shows the correlations of the voltage fluctuation measured
at a probe with the  
current transferred into a contact. As usual $p$ denotes the probe and
$\alpha$ stands for any  
of the contacts of the conductor. As pointed out at the end of subsection
\ref{langedeph}  the energy 
integrated correlations $\av{\Delta V_p\Delta I_\alpha}/V$ equal the
correlations $\av{\Delta n_p\Delta i_\alpha}$ which characterize an energy
interval in the window  
opened by the transport voltage $eV$. 

The example of the beam splitter shows that the correlation can have either
sign, depending  on whether we consider the correlation with the transmitted
($\alpha = 2$) or the reflected ($\alpha = 1$) particle current. The
dependence on the transmission probability directly reflects the shot noise
generated at the quantum point contact.  

The correlations in the MZI are independent of the phase $\Phi$. 
The correlations are a signature of incoherent processes: only carriers that
enter the probe give rise to voltage fluctuations and only carriers that leave
from the probe to the contact are correlated with the voltage at the probe.
 Note,  that the additional beam splitter $B$
the particles have to pass after leaving the probe renders the correlations
cubic in the transmission of the beam splitters (in contrast to the beam
splitter setup of the first example). Curiously, for $T_A=T_B=T$ they 
coincide with the third current cumulants \cite{lll,Forster05} of a beam
splitter with 
transmission $T$. In contrast to the simple beam splitter (first example) here
the correlation changes sign as  function of the transmission probability of
the second beam splitter $T_B$. For $T_B =1$ we have the  same situation as
for the beam splitter for $\alpha =2$, and $T_B = 0$ corresponds to the beam
splitter result for $\alpha=1$.  

Surprisingly, also the correlations for the double barrier are independent of
the phase $\Phi_d$. 
The fact that no oscillating contribution appears in the correlations is
however a feature of this particular chiral setup. For non-chiral
Aharonov-Bohm rings, for example, the current-voltage correlations do depend on
the phase $\Phi_d$ \cite{note1}.

In the triple barrier a probe is only connected to the left dot. 
The correlations of the triple barrier contain the phase $\Phi_R$ 
due to interference in the right dot but are independent of $\Phi_L$.
This is consistent with the fact that the correlation for the double barrier
is  independent of phase. 
 
All correlations shown in table \ref{table2} are independent of the coupling
strength of the probe and moreover are independent of the phase directly
adjacent to the probe. 
Further research is needed to characterize the examples where current-voltage
correlations are phase- and coupling-strength independent and the examples
where this is not the case \cite{note1}.

\section{Full counting statistics of the probe: the stochastic path integral.
}\label{FCS} 
An elegant description of the zero frequency transport is the full counting
statistics (FCS) \cite{lll,nazarov,nagaev} which permits to obtain not only the
average current and noise but the full distribution  of charges transmitted
through a conductor during a measurement time $\tau$.  For an $M$-terminal
conductor without a probe the distribution function is 
denoted by $P({\bf Q})$, where the vector quantity ${\bf
  Q}=(Q_1,Q_2,\ldots,Q_M)$ describes the charge transfered 
into each of the $M$ terminals. $P({\bf Q})$ can  be expressed in terms of 
 the cumulant generating function $S({\bf \Lambda})$,
 where  ${\bf \Lambda} =(\lambda_1,\lambda_2,\ldots
\lambda_M)$ are the conjugate variables  to ${\bf Q}$:
\begin{eqnarray}
	P({\bf Q})&=& \int d{\bf \Lambda}
	e^{S({\bf \Lambda})-i{\bf \Lambda}\cdot {\bf Q}},\label{PQ}\\
	S({\bf \Lambda})&=& \ln \sum_{\bf Q} P({\bf Q})e^{i{\bf \Lambda}\cdot
	{\bf Q}}\label{SLambda}. 
\end{eqnarray}
The sum and integrals run over all elements of the vector,$\int d{\bf
  \Lambda}=(2\pi)^{-M}\int d\lambda_1 \ldots d\lambda_M $
and  $\sum_{\bf  Q}=\sum_{Q_1\ldots Q_M}$.
Probability conservation leads to the normalization of the generating function
$S({\bf 0})=0$.

For a long measurement time $\tau$, the transmitted charge into a contact
$\alpha$ is proportional to $\tau$: $Q_\alpha=\tau I_\alpha$. Then the zero
frequency current cumulants are obtained  by taking derivatives of the
cumulant  generating function with respect to the counting variables and
evaluated  at ${\bf 
  \Lambda}=0$. The average current $\av{I_\alpha}$ and the auto- and
cross-correlations $C_{\alpha\beta}$ are given by
\begin{equation}\label{cumu}
	\av{I_\alpha}=\frac{e}{i\tau}\frac{\partial S}{\partial\lambda_\alpha},\hspace{2cm}
	C_{\alpha\beta}=
	\frac{e^2}{i^2\tau}\frac{\partial^2S}{\partial\lambda_\alpha
	\partial\lambda_\beta}. 
\end{equation}

When a probe is connected to the conductor, particles can enter and leave the
probe. However, for every particle entering the probe, a particle is re-emitted
after a delay time $\tau_d$  as described in sections \ref{langevolt} and
\ref{langedeph}. The 
functions $P({\bf Q})$ and $S({\bf \Lambda})$ have then to be determined under
the condition that no charge is 
accumulated in the probe. The transport statistics under this constraint was
developed in Refs.~\cite{short,long} using a stochastic path integral
approach. The findings are that the model 
of voltage and dephasing probes  are perfectly equivalent for the case of one
single-channel probe. For multi-channel
or multiple probes the transport statistics of voltage and dephasing probes  
differs. 

\begin{figure}
\begin{center}
\centerline{\psfig{figure=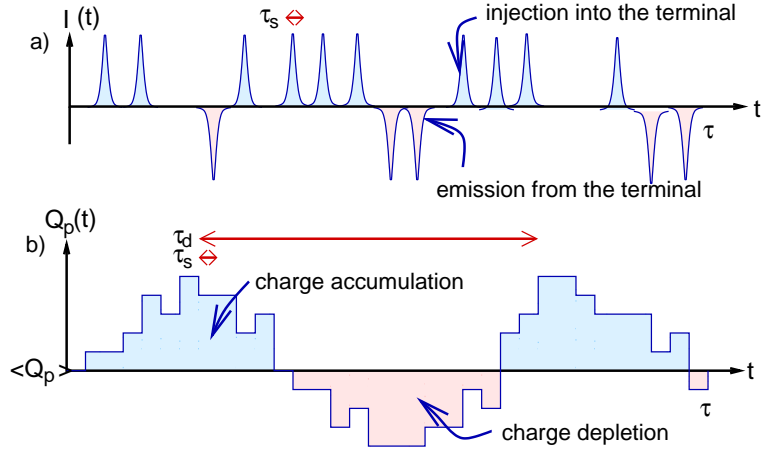,width=10.0cm}}
\caption{a) The current as a function of time at a terminal. The charge
  transfered into a terminal 
is absorbed immediately. For a one-channel contact only one charge per
time step $\tau_s=h/eV$ can enter. b) Charge on a voltage (dephasing) probe
  as a function of time. The charge on the probe accumulates and fluctuates on
  the time scale $\tau_d$, 
the delay time of the probe. The figure is for $\tau_d  \gg \tau_s$.} 
\label{chargeintime}
\end{center}
\end{figure}

Here we address the question of whether information about the statistics of the
probe itself can be obtained. Usually the quantities investigated with the FCS
are the number of charges   ${\bf Q}=\int_0^\tau dt {\bf  I}(t)$ counted
during a measurement time  $\tau$. Consequently, we now need to consider the
statistics of the  quantities $\int dt V_p(t)$ and $\int dt n_p(t)$.
 More precisely we are interested in for example 
$\frac{1}{\tau}\int dt dt' \av{V_p(t)V_p(t')}$ or  $\frac{1}{\tau}\int dt
dt'\av{n_p(t)n_p(t')}$.  
For the voltage probe we can define a normalized   phase $\phi_p$ proportional
 to the time integrated voltage  and for the dephasing   
probe a time averaged occupation number  $\bar n_p=\bar n_p(E)$, 
\begin{eqnarray}
	\phi_p&=&\frac{1}{N}\frac{e}{h}\int_0^\tau dt
	V_p(t)=\frac{1}{V\tau}\int_0^\tau dt V_p(t), \label{phi_p}\\
	\bar n_p&=&\frac{1}{\tau}\int_0^\tau dt n_p(E,t),\label{n_p}
\end{eqnarray}
where $N=eV\tau/h$.
These are the accessible quantities describing the probe, they are not numbers
of particles but phases. This is a consequence of the fact that charges
transfered into the terminals are absorbed, but the charge on the probe is
 conserved. Fig.~\ref{chargeintime} illustrates this
  difference.  
Both $\phi_p$ and $\bar n_p$ vary from measurement to measurement, because 
the injection of particles into the probe that determines
the time averaged potential and the time averaged occupation number is a
probabilistic process. 
The phase $\phi_p$ is proportional to the total charge on the probe integrated
during the measurement time $\tau$: $\phi_p=\frac{e}{CNh}\int_0^\tau dt
Q_p(t)$, with $C$ the capacitance of the voltage probe.
In Refs.~\cite{pilgram03} and \cite{SPI2} the FCS of charge
fluctuations in a chaotic cavity were investigated, conceptually this is
similar to the FCS of  $\phi_p$ addressed here. As pointed out in
Ref.~\cite{pilgram03}, the quantity $\int_0^\tau dt
Q_p(t)$ has no direct physical meaning, but can be understood as the time
spent by all electrons in the cavity (or probe) after time $\tau$, or
alternatively as the phase picked up due to the potential integrated over time
$\tau$. In analogy to this, the phase $\bar n_p$ is proportional to the charge
in an energy  interval $dE$ integrated during $\tau$. We only consider the
case that there is a large number of states in this energy interval. This
corresponds to the case of a long delay time $\tau_d$ of particles inside the
probe. 
For the stochastic path integral approach discussed here it is important that
the charge on the probe is slowly fluctuating compared to the inverse attempt
frequency $\tau_s=h/eV$: $\tau_d\gg \tau_s$.  This allows to justify the saddle
point solutions which will be used later on. A different dynamics at voltage
probes is considered in Ref.~\cite{sanjose}.

\subsection{Voltage probe.}
We define a joint probability $P({\bf Q},\phi_p)$ that ${\bf Q}$
charges are transmitted during $\tau$ and the phase due to the potential of
the voltage probe is $\phi_p$. 
By definition we have $P({\bf Q})=\int_0^1 d \phi_p P({\bf Q},\phi_p)$.
The conjugated variable  to $\phi_p$ is a number called $m_p$.  Fourier
transforms in the variables ${\bf Q}$ or $\phi_p$ or in both of them are
possible. Thus there 
are four equivalent functions containing all informations about the transport
and the probe statistics:  $P({\bf  Q},\phi_p)$, $	S({\bf
  \Lambda},m_p)$, $ 
	\xi({\bf\Lambda},\phi_p)$ and $	\zeta({\bf Q},m_p)$. Note that the
	counting fields ${\bf \Lambda}$ are periodic in $2\pi$, but $\phi_p$
	has a period of $1$. Thus the exponent in the transformation is
	multiplied by 	$2\pi$. Explicitly written we 
	have for 	example, 
\begin{eqnarray}
	P({\bf Q},\phi_p) &=& \int d {\bf\Lambda}\sum_{m_p}
	e^{S({\bf\Lambda},m_p)-i{\bf \Lambda Q}+2\pi
	i\phi_pm_p},\label{Smp}\\ 
	P({\bf Q},\phi_p) &=& \int d {\bf\Lambda}
	e^{\xi({\bf\Lambda},\phi_p)-i{\bf \Lambda Q}}. \label{xiphip} 
\end{eqnarray}
In particular we are interested in the distribution $P(\phi_p)$ of the phase
$\phi_p$ alone. It is obtained from the relation 
\begin{equation}
	P(\phi_p)=\sum_{{\bf Q}}P({\bf Q},\phi_p)=e^{\xi({\bf
	\Lambda=0},\phi_p)}.
\end{equation}
Refs.~\cite{short,long} used the stochastic path integral to calculate the
generating function of a conductor coupled to a voltage or dephasing
probe. With this formalism also the joint functions $P,S,\xi$ and $\zeta$
introduced above are accessible. A
description of the procedure is given in the appendix, here we simply
state the results.

Most directly the function $\xi({\bf\Lambda},\phi_p)$ can be determined by an
integral
\begin{equation}\label{xiint}
	e^{\xi({\bf\Lambda},\phi_p)}=\int d\lambda_p e^{\bar S_V({\bf
	\Lambda},\lambda_p,V_p=V\phi_p)}. 
\end{equation}
A very long measurement time $\tau \gg \tau_d$ defines the stationary case
considered here where the variables $\lambda_p$ and $V_p$ are
time independent. The function $\bar S_V({\bf\Lambda},\lambda_p,V_p)$ is given
by  \cite{Levitov}
\begin{equation}
\label{sv}
	\bar S_V({\bf\Lambda},\lambda_p,V_p)=\frac{\tau}{h}\int dE H_0
\end{equation}
with 
\begin{equation}
	H_0=\ln \det
	\left[1+\tilde n\left(\tilde\lambda^{\dagger}\mathcal{S}^{\dagger}
	\tilde\lambda  \mathcal{S}-1\right)\right]  \label{H0}.
\end{equation}
For a conductor with single mode contacts to the $M$ terminals and to the
probe, the  scattering matrix $\mathcal{S}$ is of dimension $(M+1)\times
(M+1)$. The matrix $\tilde n$ contains  the occupation  
numbers of the different terminals with $\tilde n=\mbox{diag}(n_1,n_2,\ldots,
n_M,n_p)$ (here $n_p\equiv f_p(V_p)$), and the
matrix $\tilde\lambda$ introduces the counting fields, 
$\tilde\lambda=\mbox{diag}(e^{i\lambda_1},e^{i\lambda_2},\ldots,
e^{i\lambda_M},e^{i\lambda_p})$.

To proceed it is useful to introduce the expressions $q_{kl}$ containing 
multi-particle scattering probabilities multiplied with the appropriate
counting fields in the contacts. The index
$l=0,1$ denotes the number of particles injected into the probe, and $k=0,1$
the number of particles emitted from the probe \cite{long}. Below, for the
examples of interest here, we give the explicite expressions for the
$q_{kl}$. With the help of the $q_{kl}$, the function $H_0$ is 
\begin{equation} 
	H_0(\lambda_p,n_p)=\ln\left[(1-n_p)\left(q_{00}+
	q_{01}e^{i\lambda_p}\right)+
	n_p\left(q_{11}+q_{10}e^{-i\lambda_p}\right)\right].\label{H0exp}
\end{equation}
Using Eq.~(\ref{H0exp}) we obtain for Eq. (\ref{sv}), with
$N=\frac{e\tau}{h}V$, 
\begin{equation}\label{SV}
	\bar S_V=N\left(\phi_p\ln[q_{11}+q_{10}e^{-i\lambda_p}]+ 
	(1-\phi_p)\ln[q_{00}+q_{01} e^{i\lambda_p} ]\right).
\end{equation}
The integral (\ref{xiint}) is solved in saddle point approximation. The saddle
 point equation $\frac{\partial \bar S_V}{\partial \lambda_p}=0$ 
 is quadratic in $e^{i\lambda_p}$.  The solution satisfying the normalization
condition $\int d\phi_p e^{\xi({\bf 0},\phi_p)}=1$ is  
\begin{equation}\label{lambdavolt}
	\hspace{-20mm}e^{i\bar
	\lambda_p}=\frac{-q_{01}q_{10}(1-2\phi_p)+
	\sqrt{(q_{01}q_{10}(1-2\phi_p))^2+
	4q_{00}q_{01}q_{10}q_{11}\phi_p(1-\phi_p)}} 
	{2q_{10}q_{11}(1-\phi_p)}.
\end{equation}
The generating function is in saddle point approximation
\begin{equation}\label{xi}
	e^{\xi({\bf\Lambda},\phi_p)}=e^{\bar S_V({\bf
	\Lambda},\bar \lambda_p,V\phi_p)}. 
\end{equation}

As demonstrated in Ref.~\cite{long,short}, the saddle point approximation is
correct when the delay time of the probe is much longer than the inverse
average attempt frequency, $\tau_d \gg h/eV$.

To obtain the cumulants of the distribution function it is important to look
at the joint generating function $S({\bf \Lambda},m_p)$. The auto- and
cross-correlations studied in section  \ref{Langevin} can be expressed in
analogy to Eqs.~(\ref{cumu}) as 
\begin{eqnarray}
	\av{\Delta V_p^2}&=&V^2\av{\Delta\phi_p^2}=-
	\frac{V^2\tau}{4\pi^2}\frac{\partial^2S}{\partial
	m_p^2},\label{auto}\\  
	\av{\Delta V_p\Delta I_\alpha}&=&\frac{V}{\tau} \av{\Delta
	\phi_p\Delta Q_\alpha}= 
	\frac{eV}{2\pi} \frac{\partial^2S}{\partial\lambda_\alpha \partial
	m_p}.\label{cross} 
\end{eqnarray}
The generating function $S({\bf \Lambda},m_p)$ is in principle obtained by a
Fourier transform of $\xi({\bf \Lambda},\phi_p)$ as indicated in
Eqs.~(\ref{Smp}) and (\ref{xiphip}). Another, more transparent way is to
express $S({\bf \Lambda},m_p)$ with the help of Eq.~(\ref{xiint}) and solve
the coupled saddle point equations. This is shown in the appendix.

\subsection{Dephasing probe.}
The distribution function for the average occupation
number $\bar n_p$ of the dephasing probe can be found in a way similar to what
is described above for the voltage probe.
Here, the distribution $P_E({\bf Q})$ is energy dependent and the vector
${\bf Q}$ denotes the charges per energy interval, ${\bf Q}={\bf Q}(E)dE$.
The function $P_E({\bf Q},\bar n_p)$ is the joint probability that 
${\bf Q}$ charges are transmitted during $\tau$ and the average occupation
number takes the value $\bar n_p$ in the energy interval $dE$. 
The conjugated variable  to $\bar n_p$ is a number called $\gamma_p$, and the
corresponding generating functions and distribution functions are 
$P_E({\bf  Q},\bar n_p)$, $	S_E({\bf	\Lambda},\gamma_p)$, $
	\xi_E({\bf\Lambda},\bar n_p)$ and $\zeta_E({\bf Q},\gamma_p)$ with for
	example 
\begin{eqnarray}
	 P_E({\bf Q},\bar n_p)&=& \int d{\bf\Lambda}
	\sum_{\gamma_p}e^{S_E({\bf\Lambda},\gamma_p)-i{\bf \Lambda
	Q}+i2\pi\bar 	n_p\gamma_p}\label{Sgammap}.  
\end{eqnarray}
With this the fluctuations and correlations of section \ref{Langevin} are
\begin{eqnarray}
	\av{\Delta
	n_p^2}&=&-\frac{dE\tau}{4\pi^2}\frac{\partial^2S}{\partial\gamma_p^2},\\  
	\av{\Delta n_p\Delta i_\alpha}&=&
	\frac{e}{2\pi} \frac{\partial^2S}{\partial\lambda_\alpha
	\partial\gamma_p}. 
\end{eqnarray}
Note that the counting fields $\lambda_\alpha$ are dimensionless and of order
$1$, but the
variable $\gamma_p$ and the generating function $S$ scale with $dE\tau/h$,
leading to the proportionality factors above. 
Here again, we are most interested in the distribution function
\begin{equation}
	P_E(\bar n_p)=	\sum_{{\bf Q}}P_E({\bf Q},\bar n_p)=e^{\xi_E({\bf
	\Lambda=0},\bar n_p)}.
\end{equation}
The function $\xi_E({\bf\Lambda},\bar n_p)$ is given from the stochastic
path integral approach in the stationary limit:
\begin{equation}\label{xiE}
	e^{\xi_E({\bf\Lambda},\bar n_p)}=\int d\lambda_p e^{\bar S_E({\bf
	\Lambda},\lambda_p,n_p=\bar n_p)}=\int d\lambda_p
	e^{\frac{dE\tau}{h}H_0(\lambda_p,\bar n_p)}. 
\end{equation}
The saddle point equation $\frac{\partial   H_0}{\partial \lambda_p}=0$ is
quadratic and the solution satisfying the normalization is 
\begin{equation}\label{lambdadeph}
	e^{i\bar \lambda_p}=+\sqrt{\frac{q_{10}\bar n_p}{q_{01}(1-\bar n_p)}}.
\end{equation}
Explicitly we obtain the generating function
\begin{equation}
	\xi_E({\bf \Lambda},\bar n_p)=\frac{dE\tau}{h}\ln\left[q_{00}(1-\bar
	n_p)+ 	q_{11}\bar n_p+2\sqrt{q_{10}q_{01}\bar n_p(1-\bar n_p)}\right].
	\label{xin_p}
\end{equation}
Even though the distribution function $P({\bf
  Q})$ of the transmitted charge only  is equal for both voltage and dephasing
  probes (in the single 
  channel case), the generating functions Eqs.~(\ref{xi}) and (\ref{xin_p})
  are different. 
Technically this originates from the different structure
  of the saddle point equations and their solutions Eqs.~(\ref{lambdadeph}) and
  (\ref{lambdavolt}).  
Physically the differences are based on the different occupation functions of
the voltage and dephasing probe, a Fermi
function and a non-equilibrium occupation function. As shown in
  Fig.~\ref{distribution}, the later is a two-step function and carries a
  probabilistic uncertainty leading to the different statistics of the number
  $\bar n_p$ compared to the phase $\phi_p$. This is
  already seen in the fluctuations $\av{\Delta V_p^2}$ and $\av{\Delta
  n_p^2}$ obtained by the Langevin approach in section \ref{Langevin}.

\section{Examples.}\label{examples}
With the formalism presented above, it is possible to discuss the full counting
statistics of voltage and occupation number fluctuations and their
correlations with transferred charge
of the examples presented in section \ref{Langevin}. Here we concentrate on the
beam splitter, the Mach-Zehnder interferometer and the double barrier.

\subsection{Beamsplitter.}\label{bsstat}
\begin{figure}
\begin{center}
	\includegraphics[width=5cm]{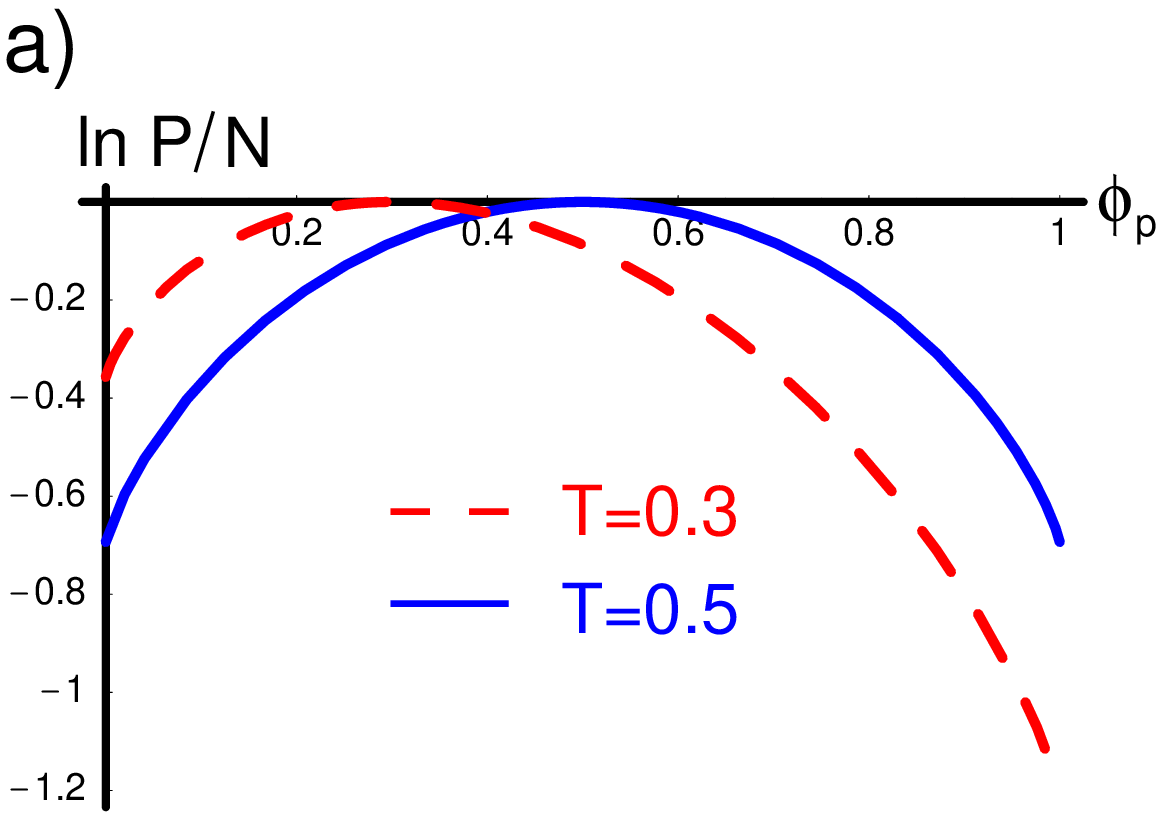} 
	\includegraphics[width=5cm]{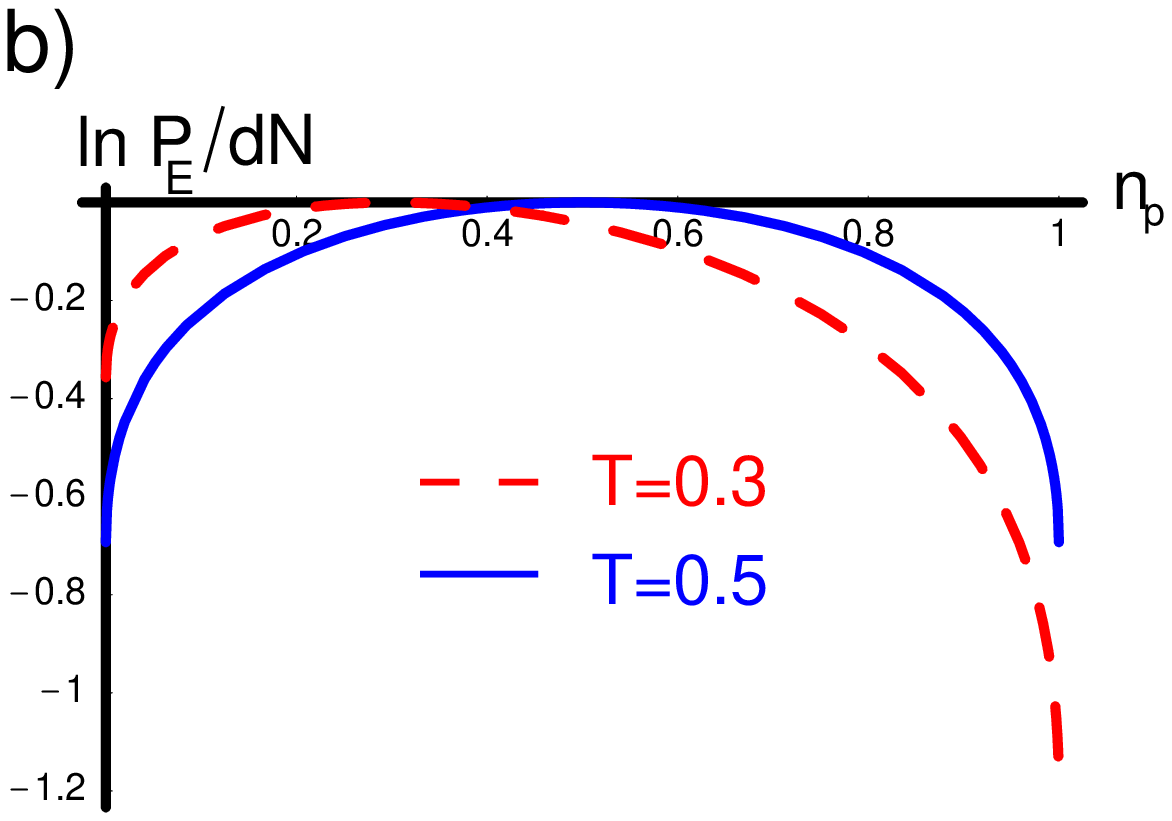} 
\caption{Distribution functions $P(\phi_p)$ of the integrated voltage (a) and
 $P_E(\bar n_p)$ of the  occupation   number (b) of the  
the probe of the beam splitter for two transmission probabilities. The
  function $P(\phi_p)$ of the integrated 
	voltage is a binomial distribution whereas the function $P_E(\bar 
	n_p)$ of the time integrated occupation number of the dephasing probe
  is 	non-binomial. The plots are normalized with respect to
	$N=\frac{eV\tau}{h}$ or $dN=\frac{dE\tau}{h}$.}\label{lnPBS} 	
\end{center}
\end{figure}

The beam splitter with transmission $T$ coupled to a probe is shown in
Fig.~\ref{beamsplitter}.
It is straightforward to evaluate the equations presented above with the
expressions for  $q_{kl}$. The $q_{kl}$ are expansion coefficients of the
determinant in Eq.~(\ref{H0}), however for this simple example they can be
found by considering the different processes in and out of the probe. $q_{00}$
describes the process that no particle enters or leaves the probe. For the
beam splitter this happens if the particle is reflected at the barrier. This
process occurs with probability $1-T$.  When the particle
passes through the barrier with probability $T$, it enters the probe, this is
$q_{01}$. The remaining $q_{10}$ and $q_{11}$ represent two-particle
processes: no particle is injected into the probe (probability $1-T$) but one
particle is emitted from the probe into terminal $2$ (counting factor
$e^{i\lambda_2}$), or one particle enters with probability $T$ and one 
 leaves the probe into terminal $2$. In conclusion, 
\begin{equation}\label{qbeam}
	\begin{array}{rclcrcl}
	q_{00}&=&1-T,&\hspace{1cm}&q_{01}&=&T,\\	
	q_{10}&=&(1-T)e^{i\lambda_2},&&
	q_{11}&=&Te^{i\lambda_2}.
	\end{array}
\end{equation}
Inserting this in Eqs.(\ref{xi}) and (\ref{xin_p}) we obtain
\begin{eqnarray}
	\xi(\lambda_2,\phi_p)&=&N
	\left(\phi_p\ln\frac{Te^{i\lambda_2}}{\phi_p}+  
	(1-\phi_p)\ln\frac{1-T}{1-\phi_p}\right), \label{xibs}\\
	\xi_E(\lambda_2,\bar n_p)&=&dN
	\ln\left[\left(\sqrt{(1-T)
	(1-\bar n_p)}+ \sqrt{Te^{i\lambda_2}\bar
	n_p}\right)^2\right],\label{xiEbs}  
\end{eqnarray}
with $N=\frac{eV\tau}{h}$ and $dN=\frac{dE\tau}{h}$.
The two functions have a very different form. The distribution functions
$P(\phi_p)=e^{\xi(0,\phi_p)}$ and $P_E(\bar n_p)=e^{\xi_E(0,\bar n_p)}$  are
plotted in Fig.~\ref{lnPBS}.  
The function $P(\phi_p)$ is a binomial distribution. 
The saddle point approximation does not lead to an binomial factor in
Eq.~(\ref{xibs}). The factor can be obtained by inserting the above expressions
Eq.~(\ref{qbeam}) into Eq.~(\ref{SV}) and evaluating the integral
Eq.~(\ref{xiint}) exactly. 
The distribution function of $\bar n_p$ for the dephasing probe
is  however non-binomial as can be clearly seen in the figure.

\begin{figure}
\begin{center}
	\includegraphics[width=6.5cm]{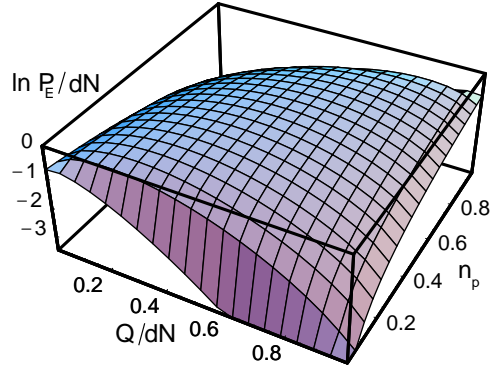} 
	\caption{The joint distribution function $P_E(Q,\bar n_p)$ of the
	charge transferred into contact $2$ and the integrated occupation
	number at the dephasing probe of the beam splitter for 
	$T=0.5$. In contrast for the voltage probe, the joint distribution
	$P(Q,\phi_p)$ is a 	one-dimensional 
	binomial function with
	$P(Q,\phi_p)=P(\phi_p)\delta_{Q,\phi_p}$. 	}\label{lnPbs3d}    
\end{center}
\end{figure}

For this simple example the joint distribution function $P(Q,\phi_p)$ can be
directly obtained with $P(Q,\phi_p)=\int \frac{d\lambda_2}{2\pi}
e^{\xi(\lambda_2,\phi_p)-i\lambda_2Q}$. We find that  
the distribution of the phase $\phi_p$ equals the distribution of the
charge transmitted into a contact of the beam splitter:
\begin{equation}\label{PQphi}
	P(Q,\phi_p)=
	e^{\xi(0,\phi_p)}\delta_{Q\phi_p}=P(\phi_p)\delta_{Q\phi_p}. 
\end{equation}
Also for the dephasing probe the joint distribution $P_E(Q,\bar n_p)$ can be
calculated (in this function $Q$ denotes the charge per energy interval).
The probability function $P_E(Q,\bar n_p)$  depends on both variables $Q$ and
$\bar n_p$, in contrast to the voltage probe, Eq.~(\ref{PQphi}) where $Q$ and
$\phi_p$ act  
effectively as one variable. This reflects the fact that the occupation
number $\bar n_p$ represents the probability for the emission of a particle,
while the 
occupation in the voltage probe  determines exactly
each scattering event. A plot of $P_E(Q,\bar n_p)$  is shown in
Fig.~\ref{lnPbs3d}. 

Although the two generating functions (\ref{xibs}) and (\ref{xiEbs}) have a
very different form, integration 
over $\phi_p$ and $\bar n_p$ respectively leads to the same generating function
for charge transmitted through a beam splitter: $\ln\int_0^1 d\phi_p
e^{\xi(\lambda_2,\phi_p)}=\int dE\ln\int_0^1 d\bar n_p
e^{\xi_E(\lambda_2,\bar n_p)}=N\ln[1+T(e^{i\lambda_2}-1)]$.
This means that a voltage or a dephasing probe, perfectly coupled to the
conductor as shown in Fig.~\ref{beamsplitter} does not
affect the statistics of transmitted charge.

\subsection{Mach-Zehnder interferometer.}\label{MZI}
\begin{figure}
\begin{center}
	\includegraphics[width=5cm]{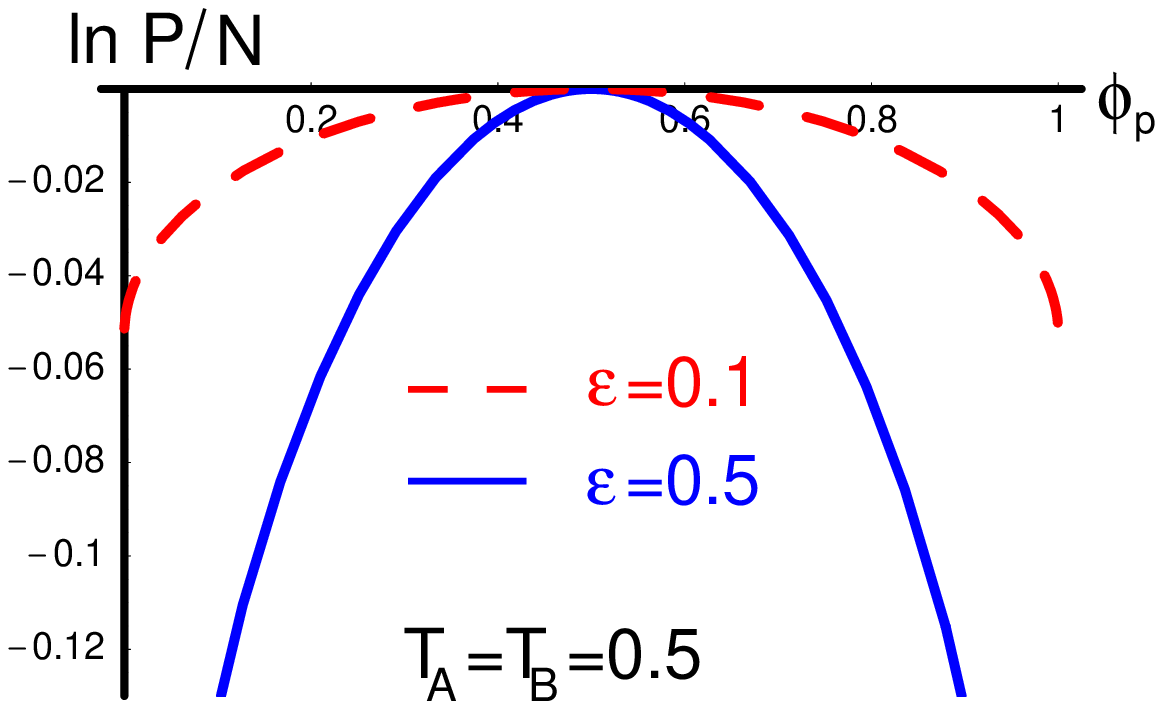} 
	\includegraphics[width=5cm]{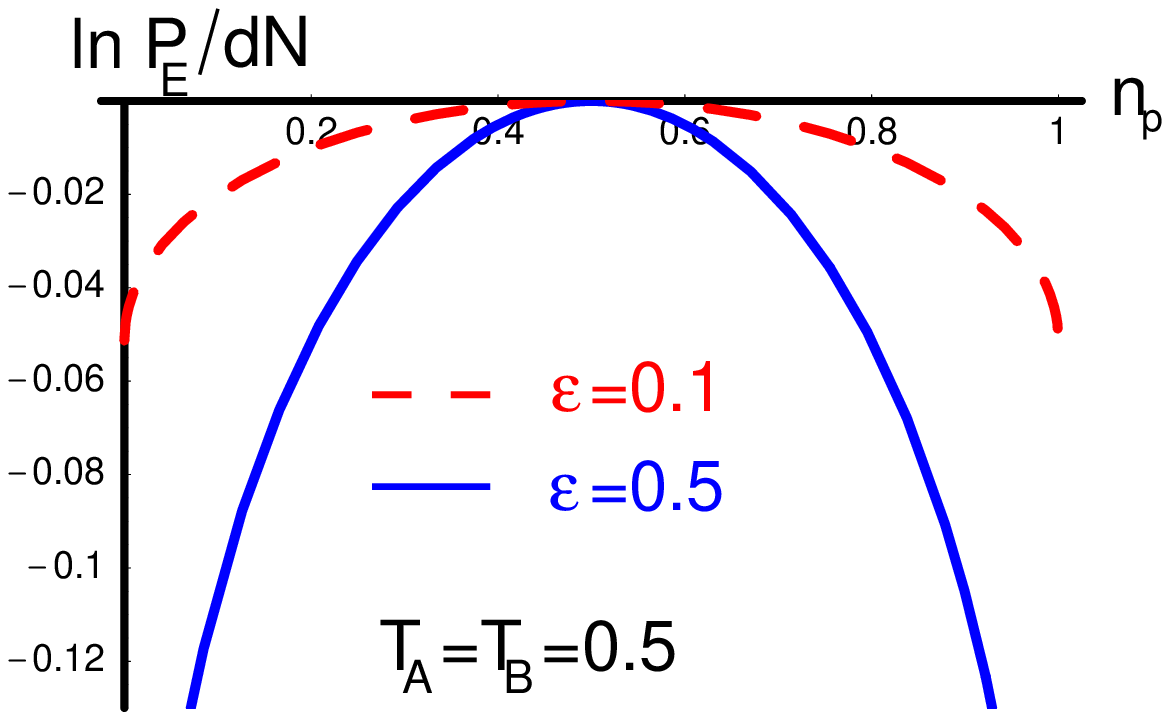} 
	\includegraphics[width=5cm]{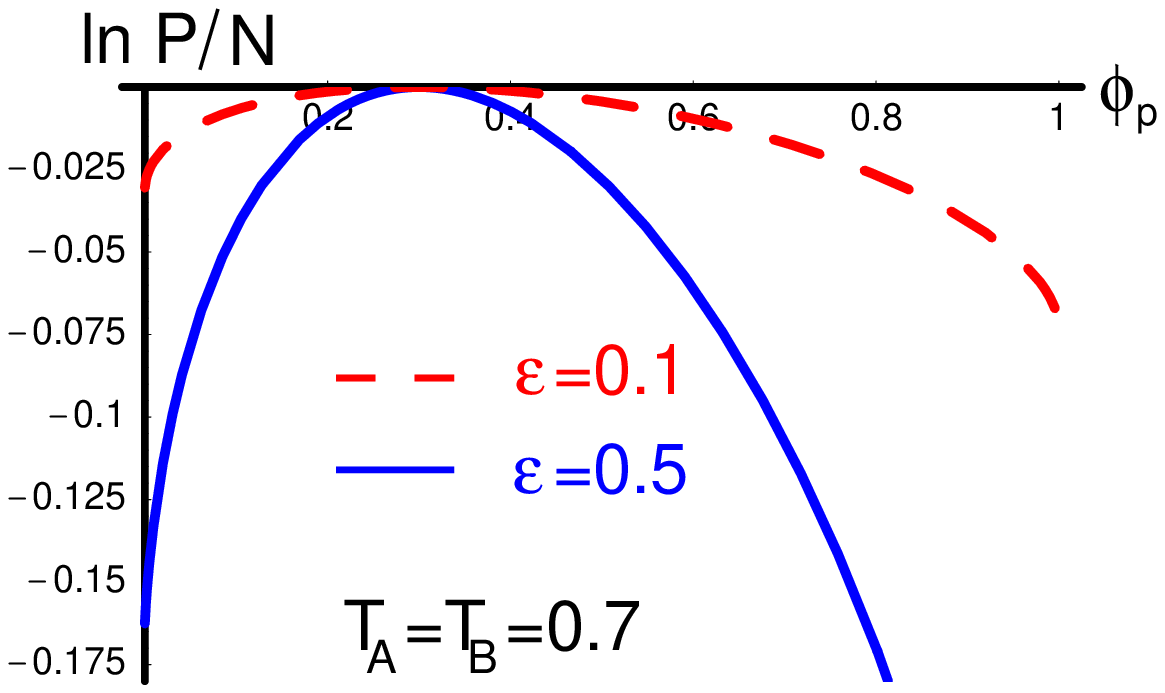} 
	\includegraphics[width=5cm]{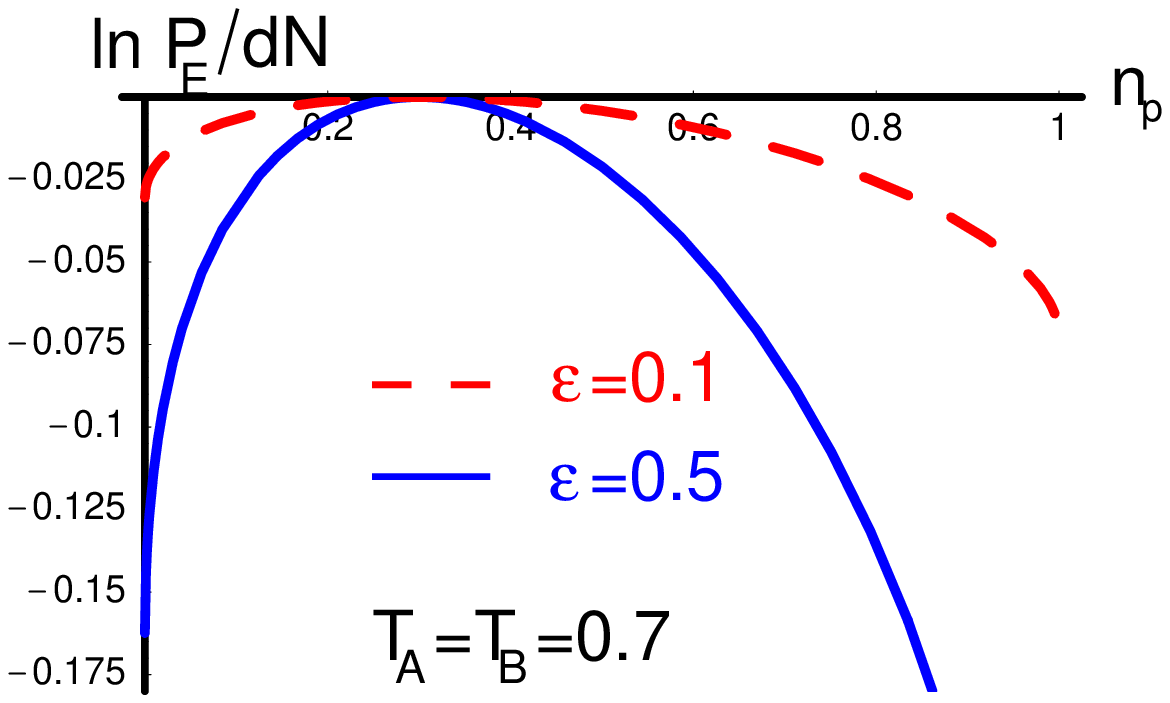} 
	\caption{The distribution functions $P(\phi_p)$ (left) of the
	integrated  
	voltage and $P_E(\bar n_p)$ (right) of the integrated occupation number
	of the probe of the Mach-Zehnder interferometer for 
different coupling strength $\varepsilon$. The case of symmetric beamsplitters
	is shown in the upper 
	figures and the case of beamsplitters with transmission $T_A = T_B =
	0.7$ is shown in the lower figures. For $\varepsilon=0.1$ both
	distribution functions look very similar (in fact they 
	are equal to first order in $\varepsilon$). For stronger coupling, 
	the tails of the distribution are broader for the dephasing
	probe. The maximum of both distribution functions lies at
	$R_A = 1-T_A$.}\label{mzi1}   
\end{center}
\end{figure}

The Mach-Zehnder interferometer is shown in Fig.~\ref{mzifig}.
The coefficients $q_{kl}$ are determined by the scattering matrix and are
derived in Ref.~\cite{long}:
\begin{eqnarray}
	q_{00}&=&\left((R_AR_B(1-\varepsilon)+T_AT_B)e^{i\lambda_3}
	+(R_AT_B(1-\varepsilon)+T_AR_B)e^{i\lambda_4}\right.\nonumber\\ 
	&&\left.+2\sqrt{R_AT_AR_BT_B(1-\varepsilon)}\cos\Phi
	(e^{i\lambda_4}-e^{i\lambda_3})\right)\\
	q_{01}&=&\varepsilon R_A\\	
	q_{10}&=&\varepsilon T_Ae^{i(\lambda_3+\lambda_4)}\\
	q_{11}&=&\left((R_AR_B+T_AT_B(1-\varepsilon))e^{i\lambda_3}
		+(R_AT_B+T_AR_B(1-\varepsilon))e^{i\lambda_4}\right.\nonumber\\
	  &&\left.+2\sqrt{R_AT_AR_BT_B(1-\varepsilon)}\cos\Phi
	(e^{i\lambda_4}-e^{i\lambda_3})\right)
\end{eqnarray}
Using these parameters, the generating functions Eqs.~(\ref{xi}) and
(\ref{xin_p}) become very long expressions, which are not explicitely written
out here. Plots of the functions at ${\bf    \Lambda=0}$ are shown in
Fig.~\ref{mzi1}. Note that both Eqs.~(\ref{xi}) and (\ref{xin_p}) contain all
information about the cumulants of the transmitted charge and about the phases
$\phi_p$ and $\bar n_p$ respectively. Thus both functions contain
terms proportional to $\cos\Phi$ due to the interference.
 However the phase-dependence $\Phi$ drops out when we study cumulants of the
 phase $\phi_p$ or the occupation number $\bar n_p$ of the probe. This has
 been already observed in section \ref{langeex} and is 
explained by the setup of the MZI:  particles enter the probe before they could
interfere and no oscillating terms appear in the cumulants of the probe.

Interestingly, the distribution functions  of $\phi_p$ and $\bar n_p$ given by
$e^{\xi(0,\phi_p)}$ and $e^{\xi_E(0,\bar n_p)}$ 
coincide to first order in the coupling parameter $\varepsilon$.  This means
that when 
only very few particles enter the probe, the distribution functions of the
potential and of the occupation function per energy respectively behave
similarly, while for a higher number of charges in the probe the differences
become more and more clear, in particular the function $P(\phi_p)$ is narrower
than $P_E(\bar n_p)$.
\subsection{Double barrier.}\label{doublebarrierstat}
\begin{figure}
\begin{center}
	\includegraphics[width=5cm]{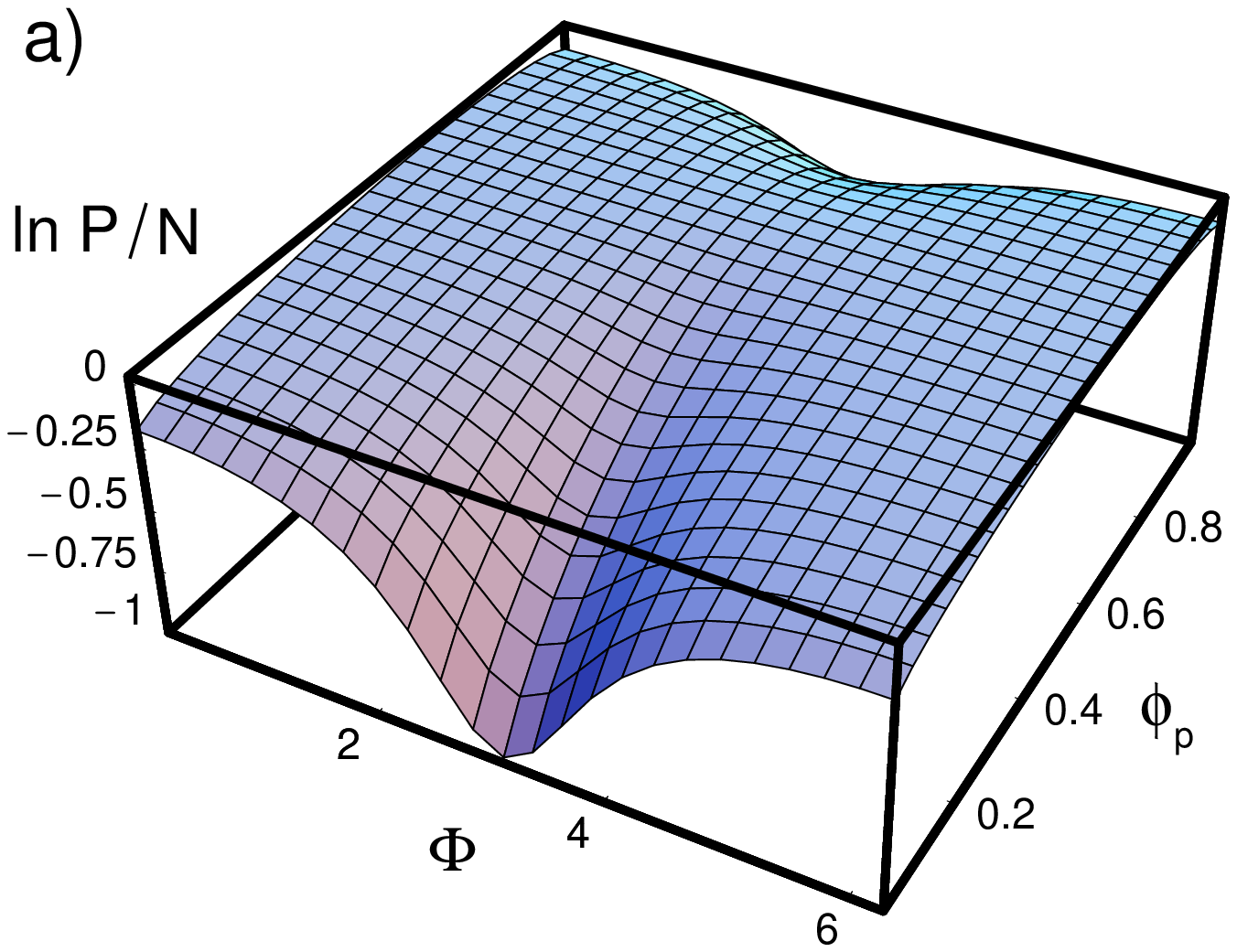} 
	\includegraphics[width=5cm]{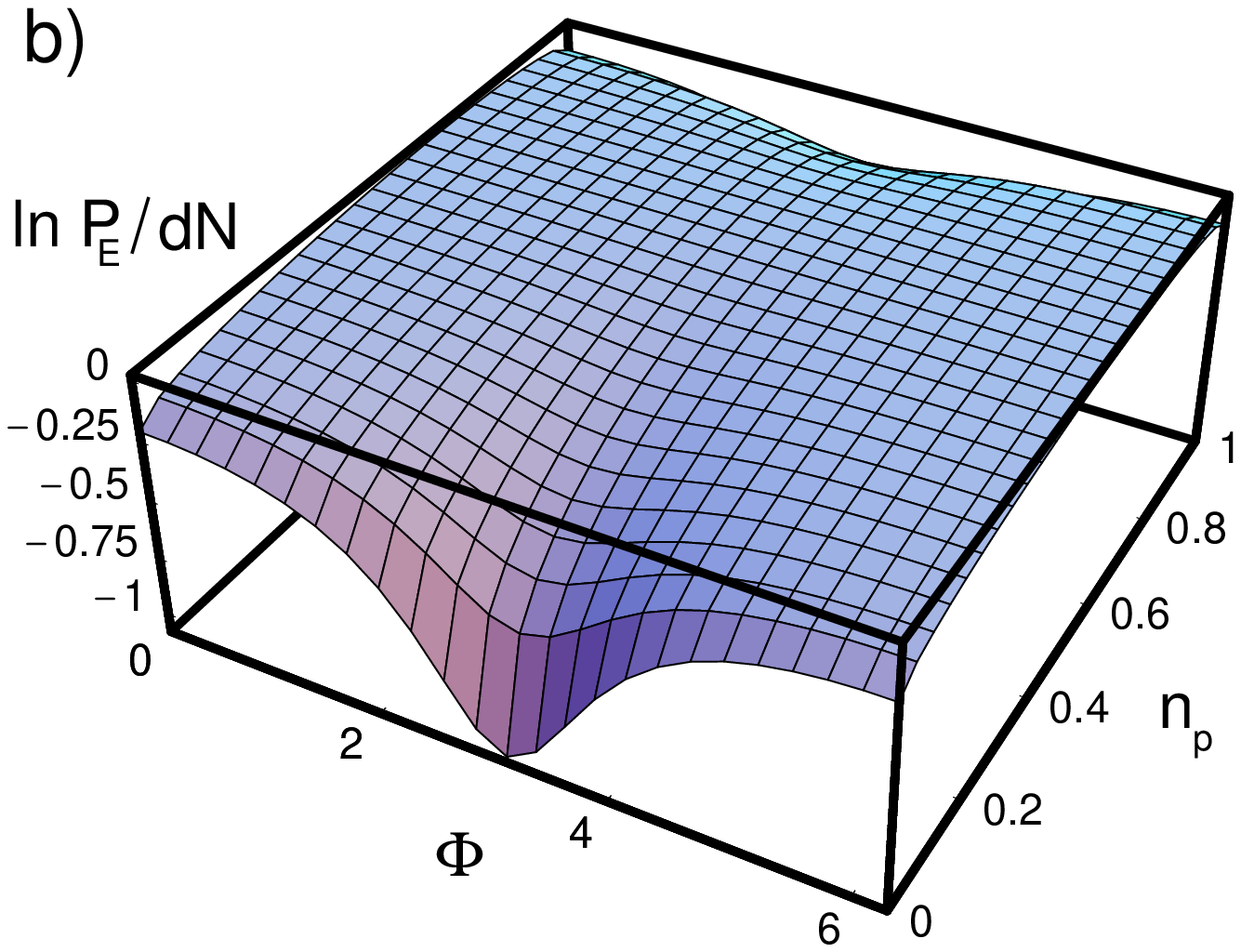} 
	\caption{The distribution function $P(\phi_p)$ for the integrated
	voltage (a) and 
	$P_E(\bar n_p)$ for the integrated occupation number (b) of the probe
	of  	the double barrier geometry 
	for  equal transmission probabilities $T_1=T_2=1/2$ and
	$\varepsilon=1-e^{-1}$ as a function of the phase $\Phi_d$.
	The different properties of the two probes are most apparent 
	around $\Phi_d = \pi$. 
	}\label{db3d}   
\end{center}
\end{figure}

The double barrier is shown in Fig.~\ref{doublebarrier}.  In this setup the
electrons enter the probe after they interfere and the distribution function
depends on the phase picked up.
The coefficients $q_{kl}$ are
\begin{eqnarray} 
	q_{00}&=&\frac{R_1+(1-\varepsilon)(R_2+T_1T_2e^{i\lambda_2})+
	2\sqrt{R_1R_2(1-\varepsilon)}\cos\Phi_d}{1+R_1R_2(1-\varepsilon)+
	2\sqrt{R_1R_2(1-\varepsilon)}\cos\Phi_d}\\    
	q_{01}&=&\frac{R_1T_2e^{i\lambda_2}\varepsilon}
	{1+R_1R_2(1-\varepsilon)+2\sqrt{R_1R_2(1-\varepsilon)}\cos\Phi_d}\\
	q_{10}&=&\frac{T_1\varepsilon}
	{1+R_1R_2(1-\varepsilon)+2\sqrt{R_1R_2(1-\varepsilon)}\cos\Phi_d}\\
	q_{11}&=&\frac{R_1(1-\varepsilon)+R_2+T_1T_2e^{i\lambda_2}+
	2\sqrt{R_1R_2(1-\varepsilon)}\cos\Phi_d}{1+R_1R_2(1-\varepsilon)+
	2\sqrt{R_1R_2(1-\varepsilon)}\cos\Phi_d}
\end{eqnarray}
Plots of the distribution functions of the voltage probe and the dephasing
probe,  Eqs.~(\ref{xi}) and (\ref{xiE}) at $\lambda_2=0$ are shown in
Fig.~\ref{db3d}.  Both functions depend on the phase  and are most
narrow at $\Phi_d=\pi$. Around this value the differences between the voltage
and dephasing probe become visible: the distribution function $P(\phi_p)$ is
narrower than $P_E(\bar n_p)$. For values further away from $\pi$ the
functions look pretty much alike. Since the transmission probability $T_{p1}$
has a maximum in $\Phi_d=\pi$, this is in agreement with our findings that
the more particles enter the probe the less broad are the distribution
functions. Higher order correlations in the double barrier are briefly
discussed in the appendix \ref{highcorr}.

\section{Conclusions.}
In this work, using the Langevin approach, we examined the auto-correlations
of voltage fluctuations at voltage probes and occupation number fluctuations
at dephasing probes. We investigated correlations between voltage (or
occupation number) fluctuations at probes and currents at a contact of the
conductor. We determined these fluctuation spectra for several
examples. Subsequently we extended this discussion to include higher order
cumulants. To achieve this we extended a stochastic path integral approach to
find the generation function for the distribution of voltage and occupation
number fluctuations and also the joint distribution of voltage (or occupation
number) fluctuations and current fluctuations. 
Interestingly we find differences between
the fluctuations of voltage and dephasing probes despite the fact that for
single channel probes the full counting statistics of the transmitted currents
is identical \cite{short,long}. This is a consequence of the fact that
occupation numbers at different energies in the dephasing probe are
uncorrelated while in a voltage probe they are correlated. The fluctuations of
the occupation number of a dephasing probe are larger than the potential
fluctuations of a voltage probe. For small coupling of the probe
-corresponding to a small number of particles entering the probe- the
differences become less important.
 It is expected that dephasing and which path detection are closely related. To
understand this we analyzed the sensitivity of the correlation functions of
voltage (or occupation number fluctuations) and currents to Aharonov-Bohm
oscillations. For the examples investigated here, we found that the
Aharonov-Bohm effect in the loop directly probed by the voltage or dephasing
contact is indeed absent. However, this is a consequence of the chirality of
the setups chosen, it is not a general feature \cite{note1}.  Further research
is needed to  systematically characterize the systems where current-voltage
correlations are independent of the Aharonov-Bohm phase.

\section*{Acknowledgment.}
This work was supported by the Swiss NSF,
the Swiss National Center of Competence in Research MaNEP,
the European Marie Curie MCTRN-CT-2003-504574 and
the Swedish VR.

\appendix
\section{Stochastic path integral.}
We show how  the Eq.~(\ref{xiint}) for the voltage probe is obtained starting
from the stochastic path integral approach \cite{SPI1,SPI2}. 
The voltage probe is described by a Fermi function with a 
potential  $V_p(t)$ fluctuating on the time scale $\tau_d$. As derived in
detail in Ref.~\cite{long}, the 
generating function $S_V({\bf \Lambda})$ (the index $V$ stands for voltage
probe) is given by the stochastic path integral
\begin{eqnarray}
	e^{S_V({\bf \Lambda})}&=& \int {\mathcal D}V_p{\mathcal
	D}\lambda_pe^{\tilde S_V({\bf \Lambda},\lambda_p,V_p)}\label{path
	integral}\\ 
	\tilde S_V&=&\frac{1}{h}\int_0^{\tau} dt \left[-i\tau_d\lambda_pe\dot
	V_p+\int  dE H_0\right].
\end{eqnarray}
The function $H_0$ is defined by Eq.~(\ref{H0}) and is determined by the
scattering matrix. The term $-i\tau_d\lambda_pe\dot V_p$ results from
the charge conservation on the probe. 
The path integral in $V_p$ and $\lambda_p$ is evaluated via a saddle point
approximation which is a good approximation for $\tau_d \gg h/eV$. For a
discussion of this important point see Refs.~\cite{short,long}. For a long
measurement time 
$\tau \gg \tau_d$, the function $\tilde S_V$ is stationary in the variables
$\lambda_p$ and $V_p$, and the saddle point equations are: 
\begin{equation}\label{Vsadpoint}
	\frac{\partial\bar S_V}{\partial\lambda_p}=0, \hspace{0.5cm}
	\frac{\partial\bar S_V}{\partial V_p}=0 \quad
	\mbox{ with } \bar S_V=\frac{\tau}{h}\int dE H_0.
\end{equation}
Starting from Eq.~(\ref{path integral}) we also find the full counting
statistics of the probe.

We define a joint probability $P({\bf Q},\phi_p)$ that ${\bf Q}$ charges are
transmitted and the phase due to the potential at the probe is $\phi_p$. 
With Eq.~(\ref{path integral}) the joint probability $P({\bf Q},\phi_p)$ is
given by 
\begin{equation}\label{Pdelta}
	\hspace{-1.2cm}P({\bf Q},\phi_p)=\int
	\mathcal{D}V_p\delta\left(\phi_p-\frac{1}{V\tau}\int_0^\tau dt 
	V_p(t)\right) \int d{\bf\Lambda}\mathcal{D}\lambda_p e^{\tilde
	S_V({\bf \Lambda},\lambda_p,V_p)-i{\bf 	\Lambda Q}}.
\end{equation}
Here the constraint $\phi_p=1/(V\tau)\int dt V_p(t)$ is introduced via a
functional delta function $\delta\left(\phi_p-\frac{1}{V\tau}\int_0^\tau dt
	V_p(t)\right)=\sum_{m_p}
	\exp\left(2\pi im_p(\phi_p-\frac{1}{V\tau}\int_0^\tau dt 
	V_p(t))\right)$.
Inserting the sum we find
\begin{equation}\label{PQphi2}
	P({\bf Q},\phi_p)=\int d{\bf \Lambda}\sum_{m_p} e^{S({\bf
	\Lambda},m_p)-i{\bf \Lambda Q}+2\pi i\phi_pm_p}
\end{equation}
where the generating function is given by
\begin{equation}\label{Smlambda}
	e^{S({\bf\Lambda},m_p)}=\int {\mathcal D}V_p{\mathcal
	D}\lambda_pe^{\tilde S_V({\bf
	\Lambda},\lambda_p,V_p)-\frac{2\pi im_p}{V\tau}\int_0^\tau dt  
	V_p(t)}. 
\end{equation}
In the stationary limit the path integrals in $\lambda_p$ and $V_p$ are
reduced to ordinary integrals. Then the sum in $m_p$ in Eq.~(\ref{PQphi2})
can be evaluated giving a delta function $\delta(\phi_p-V_p/V)$  and thus
\begin{equation}\label{PQphistat}
	\hspace{-7mm}P({\bf Q},\phi_p)=\int d{\bf \Lambda}\int d\lambda_p
	e^{\bar S_V({\bf 
	\Lambda},\lambda_p,V_p=V\phi_p)-i{\bf \Lambda Q}}=\int d{\bf \Lambda}
	e^{\xi({\bf \Lambda},\phi_p)-i{\bf \Lambda Q}}
\end{equation}
leading to Eq.~(\ref{xiint}). 

In order to obtain the joint generating function $e^{S({\bf\Lambda},m_p)}$
directly, it is convenient to start from Eq.~(\ref{Smlambda}). In the
stationary case, we have to solve the coupled saddle point equations in
$\phi_p$ and $\lambda_p$. 
Up to a constant, the solution is
\begin{eqnarray}\label{Slambdam}
	\hspace{-2.4cm}S({\bf\Lambda},m_p)&=&
	N\ln\left[q_{00}+q_{11}e^{-\frac{2\pi i
	m_p}{N}}+\sqrt{\left(q_{00}-q_{11}e^{-\frac{2\pi i
	m_p}{N}}\right)^2+4q_{01}q_{10}e^{-\frac{2\pi i
	m_p}{N}}}\right].
\end{eqnarray}
The cumulants like the auto- and cross-correlations shown in Eqs.~(\ref{auto})
and (\ref{cross})  are
then given by simple derivatives of the above function
$e^{S({\bf\Lambda},m_p)}$ taken at ${\bf \Lambda}=m_p=0$.

For the dephasing probe, the procedure outlined here is exactly the same, but
all functions concern an energy interval of width $dE$. However, the joint
generating function $S_E({\bf\Lambda},\gamma_p)$ can not be obtained exactly
unlike for the voltage probe.

\section{Higher order correlations.}\label{highcorr}

In section \ref{Langevin} we found that in all interfering structures
considered  the current-voltage correlations were independent of the phase
directly adjacent to the probe.  An interesting question is whether the full 
counting statistics discussed in sections \ref{FCS} and \ref{examples} gives
additional insight on interfering effects in 
higher order correlations.  These correlations are  joint cumulants like
$\av{\Delta V_p^k\Delta I_\alpha^l}$ which are proportional to
$\frac{d^{k+l}S({\bf \Lambda},m_p)}{dm_p^k d\lambda_\alpha^l}$ with $k,l \in
\mathbb{N}$. With the expression (\ref{Slambdam}) we can evaluate cumulants
of higher order for different setups connected to voltage probes.

For the {\bf double barrier} setup we find that $S(0,m_p)$ which
determines the distribution function of the time-integrated voltage $P(\phi_p)$
depends on the phase in the dot
$\phi_d$ as already seen in section \ref{doublebarrierstat}.
All cumulants linear in $\Delta V_p$ but to any order in the current in lead
$2$  are independent of the 
phase $\phi_d$ and the coupling strength $\varepsilon$ (the proportionality
sign hides factors of $2\pi i$):
\begin{equation}
	\hspace{-2cm}\left.\frac{\partial
	S(\lambda_2,m_p)}{\partial m_p}\right|_{m_p=0}\propto 
	\frac{e^{i\lambda_2}(1-R)^2+4R+ 
	\sqrt{e^{2i\lambda_2}(1-R)^4+4e^{i\lambda_2}R(1-R)^2}} 
	{2(e^{i\lambda_2}(1-R)^2+4R)} .
\end{equation}
The above function at $\lambda_2=0$ determines the mean voltage $\bar V_p$,
the first derivative gives the current-voltage correlations that were already
obtained in section \ref{langeex} etc. 
Cumulants quadratic in $\Delta V_p$ however always contain a contribution
proportional to $\cos\phi_d$,
\begin{equation}
	\hspace{-2cm}\left.\frac{\partial^2S(\lambda_2,m_p)}{\partial
	m_p^2}\right|_{m_p=0}\propto  
	\frac{R((e^{i\lambda_2}(1-R)^2+2R)(2-\varepsilon)+4R
	\sqrt{1-\varepsilon}\cos\Phi_d)} 
	{N\varepsilon(1-R)e^{i\lambda_2/2}(e^{i\lambda_2}(1-R)^2+4R)^{3/2}}.
\end{equation}
For the {\bf Mach-Zehnder interferometer}, the function $S(0,m_p)$
depends only on the transmission probability of the first beam splitter, $T_A$,
and the coupling parameter $\varepsilon$ in agreement with our findings of
sections \ref{langeex} and \ref{MZI}. Also for this example the
cumulants linear in $\Delta V_p$ but to any order in current do not
depend on the Aharonov-Bohm phase $\Phi$ or on the coupling strength, but on
both $T_A$ and $T_B$. We omit to print the whole expression which is lengthy.  
Remarkably, higher order correlations contain interfering terms, for example
\[
	\hspace{-2.5cm}\frac{\partial^3S({\bf
	\Lambda},m_p)}{\partial
	m_p^2\partial\lambda_3}\propto\frac{R_AT_A}{N\varepsilon} 
	\left((R_A-T_A)(R_B-T_B)(2-\varepsilon)+
	4\sqrt{R_AT_AR_BT_B(1-\varepsilon)} \cos\Phi\right).
\]
This is an example how full counting statistics can give additional
information which is not present in the auto- and cross correlations.

\end{document}